\documentclass[superscriptaddress,dvipsnames,nofootinbib,amsmath,amssymb,prd,notitlepage,twocolumn]{revtex4-1}
\usepackage{color}
\usepackage{bm}
\usepackage{graphicx}
\usepackage{amsmath}
\usepackage{amssymb}
\usepackage{enumitem}
\usepackage{hyperref}
\usepackage{subfigure}
\usepackage{array}
\usepackage{multirow}
\usepackage[english]{babel}
\usepackage{tensor}
\usepackage{comment}
\usepackage[normalem]{ulem}
\usepackage[dvipsnames]{xcolor}

\def\be{\begin{equation}}
\def\ee{\end{equation}}
\def\ba{\begin{eqnarray}}
\def\ea{\end{eqnarray}}
\def\l{\left}
\def\r{\right}

\allowdisplaybreaks

\def\l{\left}
\def\r{\right}

\def\be{\begin{equation}}
\def\ee{\end{equation}}
\def\ba{\begin{eqnarray}}
\def\ea{\end{eqnarray}}

\newcommand{\bea}{\begin{eqnarray}}
\newcommand{\eea}{\end{eqnarray}}
\newcommand{\dv}{\ensuremath{\mathrm{d}}}

\newcommand{\al}{\ensuremath{\alpha}}

\newcommand{\ka}{\ensuremath{\kappa}}

\begin{document}

\title{Inverse non-metricity in $f(Q)$ gravity: cosmology and observational constraints}

 \author{Lu\'is Atayde}
 \email{luisbbatayde@gmail.com}
 \author{Simão Marques Nunes}
 \email{simaomnunes@gmail.com}
\affiliation{Instituto de Astrof\'{i}sica e Ci\^{e}ncias do Espa\c{c}o, Faculdade de Ci\^{e}ncias da Universidade de Lisboa, Edif\'{i}cio C8, Campo Grande, P-1749-016 Lisbon, Portugal}
 \affiliation{Departamento de F\'{i}sica, Faculdade de Ci\^{e}ncias da Universidade de Lisboa, Edif\'{i}cio C8, Campo Grande, P-1749-016 Lisbon, Portugal}

\author{Noemi Frusciante}
\email{noemi.frusciante@unina.it}
\affiliation{Dipartimento di Fisica ``E. Pancini", Universit\`a degli Studi  di Napoli  ``Federico II", Compl. Univ. di Monte S. Angelo, Edificio G, Via Cinthia, I-80126, Napoli, Italy}
\affiliation{INFN Sezione di Napoli, Università degli Studi di Napoli “Federico II”,\\Compl. Univ. di Monte S. Angelo, Edificio G, Via Cinthia, I-80126, Napoli, Italy}

\begin{abstract}
We study a minimal modified gravity scenario in the symmetric teleparallel (non-metricity) formulation, focusing on an inverse non-metricity term with $f(Q)=Q+M^4 Q^{-1}$. The model does not introduce additional free parameters relative to $\Lambda$CDM, but modifies the late-time expansion and linear growth via an enhanced effective gravitational coupling. We identify key signatures: an enhanced matter power spectrum and CMB lensing, alongside a reduced late-time ISW effect and a shift in CMB peak positions. We confront the model with CMB data alone and in combination with BAO, RSD, SNIa, and DES large-scale structure data, considering both fixed minimal neutrino mass and varying $\Sigma m_\nu$. We find that the model typically prefers higher $H_0$ than $\Lambda$CDM, alleviating the $H_0$ tension, while its boosted growth tends to increase clustering amplitudes unless offset by larger neutrino masses when $\Sigma m_\nu$ is free. Overall, CMB-only data provide at most weak statistical support compared to $\Lambda$CDM, whereas late-time measurements impose tight restrictions that largely remove any improvement, positioning this model as a minimal yet strongly constrained alternative to dark energy.
\end{abstract}

\maketitle
\date{\today}

\section{Introduction}

The standard $\Lambda$CDM cosmological model, embedded in General Relativity (GR), has achieved impressive success across a wide range of probes, notably the cosmic microwave background (CMB), baryon acoustic oscillations (BAO) and large scale structures (LSS)~\cite{Planck:2018vyg}. Nonetheless, $\Lambda$CDM remains a phenomenological description with persistent problems, most prominently the cosmological constant ($\Lambda$) and coincidence problems \cite{Weinberg:2000yb,Velten:2014nra,Bull:2015stt}, and observational discrepancies~\cite{CosmoVerseNetwork:2025alb}. While the Hubble constant ($H_0$) tension between early- and late-time inferences is the most significant issue, the LSS amplitude discrepancy, often expressed in terms of $S_8$, appears to be lessening in recent combined analyses and modeling efforts~\cite{Wright:2025xka}.

A particularly notable recent development is a mild-to-moderate preference for dynamical dark energy in $w_0$--$w_a$ parameterizations, with $w_0>-1$ and $w_a<0$ favored at the $\mathcal{O}(2.5$--$4)\sigma$ level depending on the dataset combination~\cite{DESI:2024mwx}. Follow-up analyses broadly agree that the data are compatible with (and in some cases prefer) time variation in the dark-energy equation of state, while remaining consistent with $\Lambda$CDM within $\lesssim 2\sigma$ for several model and data choices~\cite{Giare:2024rgi,Dinda:2024aag}.

Against this backdrop, modified gravity (MG) offers a geometric route to late-time acceleration that goes beyond a constant vacuum term. Within the geometrical trinity of gravity, GR can be reformulated equivalently in terms of curvature, torsion, or non-metricity. The symmetric teleparallel formulation attributes gravity to the non-metricity scalar $Q$ with vanishing curvature and torsion. GR is recovered by the linear Lagrangian $f(Q)=Q$ (often dubbed STEGR or ``coincident GR'')~\cite{Jimenez:2017tkx}. Nonlinear extensions to general functions $f(Q)$ yield to field equations with rich cosmological phenomenology~\cite{BeltranJimenez:2019tme,Heisenberg:2023uxr,Frusciante:2021sio,Albuquerque:2022eac,Goncalves:2024sem}, and have already been confronted with a variety of data at background and perturbative levels~\cite{Lazkoz:2019sji,Atayde:2023aoj,Atayde:2021pgb,Ayuso:2020dcu,Barros:2020bgg,Boiza:2025xpn,Nashed:2025usa,Sahlu:2024pxq,Wang:2024eai,Shi:2023kvu,Ferreira:2023awf,Mandal:2021bpd,Koussour:2022zgo,Ferreira:2022jcd,DAgostino:2022tdk,ElBourakadi:2023sch,Narawade:2023rip,Anagnostopoulos:2021ydo,Sokoliuk:2023ccw,Anagnostopoulos:2022gej,Aggarwal:2022eae,Koussour:2023rly,Najera:2023wcw,Oliveros:2023mwl,Sakr:2024eee,YADAV2024114,Pradhan:2024eew,Sharma:2024gbl,Oliveira:2025qvn,2025EPJC...85..656K,Sultanaa:2025ooz,Karmakar:2025yng,Li:2025msm}. 

In this work, we investigate the inverse non-metricity model,  where  $f(Q)=Q+M^4 Q^{-1}$ \cite{BeltranJimenez:2019tme}
 and is the natural analogue of inverse-curvature scenarios in $f(R)$ gravity \cite{Carroll:2003wy}. We develop the background dynamics and investigate the linear perturbation theory, and confront the model with current cosmological datasets CMB, BAO, SNe~Ia, redshift-space distortions (RSD), galaxy clustering and cosmic shear. Our aim is to assess whether inverse $f(Q)$ gravity can replicate the observed expansion history and structure growth, and to quantify its viable parameter space in light of cosmological tensions.

Additionally, we also investigate the impact of massive neutrinos in this scenario. Massive neutrinos imprint distinctive signatures in the CMB and in late-time LSS \cite{Lesgourgues:2006nd} that can partially overlap with, and therefore be degenerate with, the effects induced by MG theories \cite{Barreira:2014ija,Shim:2014uta,Baldi:2013iza,He:2013qha,Dossett:2014oia,Hojjati:2011ix,Motohashi:2012wc,Hu:2014sea,Bellomo:2016xhl,Frusciante:2020gkx,Ballardini:2020iws,Atayde:2023aoj}. In linear theory, neutrino free-streaming suppresses the growth of matter perturbations below a characteristic scale, reducing the small scale matter power spectrum, while in the CMB it affects the late-time gravitational potentials (and hence the large-angle anisotropies) and decreases the amount of CMB lensing, with smaller secondary smoothing of the acoustic peaks. Because several MG scenarios can alter the expansion history and the growth/lensing sector in ways that partially mimic (or compensate) neutrino mass effects, the inferred bounds on $\Sigma m_\nu$ are model dependent. In the minimal $\Lambda$CDM+$\Sigma  m_\nu$ framework, the combination of \textit{Planck}~2018 CMB data with BAO measurements yields a stringent upper limit $\Sigma m_\nu \lesssim 0.12~\mathrm{eV}$ (95\% C.L.)\,\cite{Planck:2018vyg} and $\Sigma m_\nu < 0.14~\mathrm{eV}$ (95\% C.L.)
 combining DESY6 3x2pt
 with CMB, and low-z datasets
 such as DESI DR2 BAO, DES SNIa and SPT clusters \cite{DES:2026fyc}.
 In the context of symmetric teleparallel $f(Q)$ gravity, it has been found a weaker bound driven by degeneracies with the MG parameter, obtaining $\Sigma m_\nu < 0.277~\mathrm{eV}$ (95\% C.L.) for the combination of CMB, BAO, RSD and SNIa \,\cite{Atayde:2023aoj}.

The paper is organized as follows. Section~\ref{sec_theory} reviews $f(Q)$ gravity and derives the background and linear perturbation equations. It also presents the inverse non-metricity $f(Q)$ model and discusses its background solutions and MG coupling at linear scales. Section~\ref{sec_theoretical_predictions} discusses the theoretical predictions for the background expansion history and on cosmological power spectra. Section~\ref{sec_Data} presents the data sets and the analysis methodology, and Section~\ref{sec_cosmological_constraints}  discusses the cosmological constraints. We conclude in Section~\ref{sec_conclusion}.

\section{$f(Q)$-theory}\label{sec_theory}

Within the metric--affine formulation of gravity, where the metric $g_{\mu\nu}$ and the affine connection $\Gamma^{\alpha}{}_{\mu\nu}$ are treated as independent dynamical variables, the action for $f(Q)$ gravity can be expressed as~\cite{BeltranJimenez:2018vdo, BeltranJimenez:2017tkd, BeltranJimenez:2019tme}
\begin{equation}
    \label{eq:action}
    \mathcal{S} = \int d^4x \, \sqrt{-g} \left[-\frac{1}{2\kappa^2} f(Q) + \mathcal{L}_{\rm m}(g_{\mu\nu}, \chi_i)\right],
\end{equation}
where $g$ denotes the determinant of the metric, $\kappa^2 = 8\pi G_N / c^4$ with $G_N$ being the Newtonian constant and $c$ the speed of light, while $\mathcal{L}_{\rm m}$ is the matter Lagrangian density depending on the metric and matter fields $\chi_i$. The function $f(Q)$ is an arbitrary function of the non-metricity scalar $Q = - Q_{\alpha\mu\nu} P^{\alpha\mu\nu}$, where the non-metricity tensor is defined as $Q_{\alpha\mu\nu} \equiv \nabla_{\alpha} g_{\mu\nu}$ and the covariant derivative $\nabla_\alpha$ is constructed from $\Gamma^{\alpha}{}_{\mu\nu}$.
The tensor $P^{\alpha}{}_{\mu\nu}$ is given by
\begin{equation}
    P^{\alpha}{}_{\mu\nu} = -\frac{1}{2} L^{\alpha}{}_{\mu\nu} 
    + \frac{1}{4}(Q^{\alpha} - \tilde{Q}^{\alpha}) g_{\mu\nu} 
    - \frac{1}{4} \delta^{\alpha}_{(\mu} Q_{\nu)},
\end{equation}
where  $L^{\alpha}{}_{\mu\nu} = \frac{1}{2}(Q^{\alpha}{}_{\mu\nu} - Q_{(\mu\nu)}{}^{\alpha})$, and  $Q_{\alpha} = g^{\mu\nu} Q_{\alpha\mu\nu}$ and $\tilde{Q}_{\alpha} = g^{\mu\nu} Q_{\mu\alpha\nu}$.

Variation of the action~\eqref{eq:action} with respect to the metric yields the field equations~\cite{BeltranJimenez:2017tkd, BeltranJimenez:2019tme, Dialektopoulos:2019mtr, Anagnostopoulos:2021ydo}:
\begin{equation}
    \label{eq:field-eqs}
    \frac{2}{\sqrt{-g}} \nabla_{\alpha}\!\left(\sqrt{-g} f_Q P^{\alpha\mu}{}_{\nu}\right)
    + \frac{1}{2} \delta^{\mu}{}_{\nu} f
    + f_Q P^{\mu\alpha\beta} Q_{\nu\alpha\beta}
    = T^{\mu}{}_{\nu},
\end{equation}
where $f_Q \equiv \partial f / \partial Q$, $\delta^{\mu}{}_{\nu}$ is the Kronecker delta, and $T_{\mu\nu}$ is the energy--momentum tensor of matter, which we assume to have the perfect-fluid form $T^{\mu}{}_{\nu} = \mathrm{diag}(-\rho, p, p, p)$, with $\rho$ and $p$ denoting the energy density and isotropic pressure, respectively.

Variation with respect to the affine connection gives the second set of field equations~\cite{BeltranJimenez:2017tkd, BeltranJimenez:2019tme}:
\begin{equation}
    \label{eq:connection-eqs}
    \nabla_{\mu} \nabla_{\nu}\!\left(\sqrt{-g} f_Q P^{\mu\nu}{}_{\alpha}\right) = 0,
\end{equation}
which holds when the hypermomentum vanishes.

In this paper, without loss of generality~\cite{BeltranJimenez:2018vdo}, we  adopt the coincident gauge, i.e. a coordinate choice such that ${\Gamma^\al_{\phantom{\al}\mu\nu}=0}$, \cite{BeltranJimenez:2017tkd, BeltranJimenez:2019tme}.

\subsection{Background equations}

In the following, we focus on the cosmological background dynamics by applying the above field equations to a homogeneous and isotropic universe, characterized by the spatially flat  Friedmann-Lema{\^i}tre-Robertson-Walker (FLRW)  line element
\be
    \label{eq:FLRW-flat}
    \dv s^2=-\dv t^2+a(t)^2\dv x^i \dv x_i\,,
\ee
where $a(t)$ is the scale factor, $t$ is the cosmic time, and $i$ runs over the spatial coordinates. 
Using the FLRW metric the non-metricity scalar is $Q = 6H^2 $, where $H\equiv\dot a/a$ defines the Hubble function. Throughout this work overdots denote derivatives with respect to $t$. 

The matter sector is modeled as a collection of non-interacting perfect fluids, each characterized by an energy density $\rho_i$ and pressure $p_i$. The total energy density and pressure are given by $\rho \equiv \Sigma_i \rho_i$ and $p \equiv \Sigma_i p_i$, respectively. Each fluid component is individually conserved, satisfying the continuity equation
\begin{equation}
    \label{eq:Continuity}
    \dot{\rho}_i +3H\left(\rho_i + p_i\right)=0,
\end{equation}
which leads to the standard scaling with the scale factor $a$,
\begin{equation}
    \label{eq:densities}
    \rho_i = \rho_{i}^0\, a^{-3(1+w_i)},
\end{equation}
assuming a constant equation of state parameter $w_i \equiv p_i / \rho_i$ for each species (specifically, $w_{c,b}=0$ for cold dark matter and baryons ( hereafter $m=c+b$), and $w_r=1/3$ for radiation ($r$)). Here and in the following, the sub/super-script $0$ denotes present-day values. We further define the usual density parameters as $\Omega_i \equiv \kappa^2 \rho_i / (3H^2)$. 

The modified Friedmann  equations follow from Eqs. \eqref{eq:field-eqs}\eqref{eq:connection-eqs} and are:
\begin{eqnarray}
    \label{eq:Fried1}
    &&6f_QH^2-\frac12f = \ka^2\rho \,,\\
    \label{eq:Fried2}
    &&\l(12H^2f_{QQ}+f_Q\r)\dot{H}=-\frac{\ka^2}{2} \l(\rho+p\r)\,.  
\end{eqnarray}

We rewrite Eqs.~\eqref{eq:Fried1} and~\eqref{eq:Fried2} similarly to the ones we are used to in $\Lambda$CDM, where the density of $\Lambda$ is replaced by the one of a `dark $Q$--fluid':
\bea
    \label{eq:Fried-fQ}
    &&3H^2=\ka^2(\rho+\rho_Q) \,,\\
    \label{eq:Raych-fQ}
    &&\dot{H}=-\frac{\ka^2}{2}\l(\rho+p+\rho_Q+p_Q\r) \,,
\eea
with the density and the pressure of the dark fluid defined as
\bea
    \label{eq:rhoQ}
    \rho_Q&=&\frac{1}{\ka^2}\l(3H^2+\frac{1}{2}f-6f_QH^2\r) \,,\\
    \label{eq:pQ}
    p_Q&=&\frac{1}{\ka^2}\l[2\dot{H}(12H^2f_{QQ}+f_Q-1)\r.\nonumber \\
    &+&\l.H^2(6f_Q-3)-\frac{1}{2}f\r] \,.
\eea

The resulting equations determine an expansion history that generally departs from the standard $\Lambda$CDM one, depending on the specific functional form of $f(Q)$. Only for the particular choice $f(Q) = Q + M\sqrt{Q}$ \cite{BeltranJimenez:2019tme}, with $M$ being a constant parameter, the background evolution exactly reproduces the $\Lambda$CDM expansion history.

\subsection{Linear cosmological scales }\label{sec_linear_cosmological_scales}

In order to study the evolution of cosmic structures, we now consider linear cosmological scalar perturbations around the homogeneous and isotropic FLRW background.
The corresponding perturbed field equations in $f(Q)$ gravity are lengthy and can be found in full detail in Ref.~\cite{BeltranJimenez:2019tme}. For the purpose of this work, we adopt the quasi-static approximation, valid on sub-horizon scales, under which the Poisson equation in Fourier space reads~\cite{BeltranJimenez:2019tme}:
\begin{equation}
\label{eq:Poisson}
 -\frac{k^2}{a^2} \Psi = \frac{\kappa^2}{2}\, \frac{1}{f_Q}  \rho_\mathrm{m} \delta_{\rm m}\, , 
\end{equation}
where, as usual, $\Psi$ is the gravitational potential,  $\delta_{\rm m}$ is the matter density perturbation and $k$ is the wave number.  Additionally,   in this approximation, from the space equations in presence of standard matter sources and in the absence of anisotropic stress, it follows that the two gravitational potentials of the metric $\Psi$ and $\Phi$, are the same. Finally, under this approximation the  continuity equation for the matter field assumes formally the same form of $\Lambda$CDM, but modifications enter through the  Poisson equation
\begin{equation}
\ddot{\delta}_{\rm m}+2H\dot{\delta}_{\rm m}+\frac{k^2}{a^2}\Psi=0\,.
\end{equation}

Similarly to what it is done for other MG theories~\cite{Amendola:2007rr,Silvestri:2013ne} we can also define an effective  gravitational coupling,  which governs the modifications of gravity on the clustering of matter:
\begin{equation} \label{mu fQ}
   \mu(a)=\frac{1}{f_Q}\,.
\end{equation}
This provides information on the  strength of the gravitational interaction with respect to the standard model.

\subsection{The inverse non-metricity $f(Q)$ model}

 The functional form of $f(Q)$ that we select is an inverse power of the non-metricity scalar, hence we call it the inverse non-metricity model.  We  adopt \cite{BeltranJimenez:2019tme}:
\begin{equation}
f(Q)=Q+\frac{M^4}{Q}\, , 
\label{eq:fq model 2}
\end{equation}
where $M$ is a constant with a dimension of a mass. It has been found that this form of $f(Q)$ is relevant for dark energy applications  because it allows for solutions that naturally give a transition from a matter dominated Universe to an accelerating Sitter Universe \cite{BeltranJimenez:2019tme} and  introduces modifications in both the expansion history and  perturbations.

At background level, from eqs. \eqref{eq:rhoQ} and \eqref{eq:pQ}, the inverse non-metricity model has a density and a pressure for the $Q$ field that assume the forms:
\begin{eqnarray}   
&&\rho_{Q}=\frac{M^4}{4\kappa^2  H^2}\, ,
\label{eq:rho_model_2}\\
&&p_{Q}=-\frac{M^4}{12 \kappa^2 H^2} \left(3 -2 \frac{\dot{H}}{H^2}\right)\, ,
\label{eq:p_model_2}
\end{eqnarray}
and the  density parameter is
\be
\Omega_Q= \frac{M^4}{12  H^4} \,.
\label{eq:Omega_Q}
\ee
At present time,  the above expression simply reads $\Omega_Q^0=M^4/12H_0^4$. 
The latter can be used to write the $M$ parameter in terms of $\Omega_Q^0$ as follows: 
\be
M^4=12 \Omega_Q^0 H_0^4 \, .
\label{eq:defenition_M}
\ee
Additionally, the flatness condition ($\Sigma_i \Omega_i=1$), when evaluated at present time, allows us to write $\Omega_Q^0$ in terms of the other density parameters, i.e. $\Omega_Q^0=1-\Omega_{\rm m}^0 -\Omega_r^0$.
Therefore, the inverse non-metricity model does not have additional parameters with respect to $\Lambda$CDM.

From Sec. \ref{sec_linear_cosmological_scales}, we find that linear perturbation theory is modified with respect to $\Lambda$CDM by the function:
\begin{equation} 
    \mu(a)=\left[1-\frac{ \Omega_{Q}^0}{3}\left(\frac{H_0}{H(a)}\right)^4\right]^{-1}\,,
    \label{eq:mu_model2}
\end{equation}
which suggests a stronger gravitational interaction with respect to $\Lambda$CDM.

\section{Theoretical predictions}\label{sec_theoretical_predictions}

In order to provide theoretical predictions for the inverse non-metricity model, we have implemented it within the Einstein–Boltzmann solver \texttt{MGCAMB} \cite{Hojjati:2011ix,Zucca:2019xhg}. This new patch allows for the consistent evolution of both the modified background expansion and the linear perturbations according to the model’s dynamics. With this implementation, \texttt{MGCAMB} can now compute cosmological observables that fully account for the effects of the inverse non-metricity model on both the background and the growth of structures.

To understand the effects created by our models we are going to fix the values of the following parameters:
$\Omega_c  h^2=0.12011$,
$\Omega_b  h^2=0.022383$,
$H_0=67.32$ km/s/Mpc,
$A_s= 2.1 \times 10^{-9}$ and
$n_s=0.96605$, and for massive neutrinos we consider $\Omega_\nu h^2=0.00064$ with $\Sigma m_\nu=0.06$ eV; $\Omega_\nu h^2=0.0054$ with $\Sigma m_\nu=0.50$ eV and $\Omega_\nu h^2=0.0097$ with $\Sigma m_\nu=0.90$, where $\Omega_\nu$ is  the energy density parameter of neutrinos,  $h=H_0/100$, $A_s$ initial amplitude of primordial perturbations  and  $n_s$ is scalar spectral index.

\begin{figure}[t]
\centering
\includegraphics[width=0.43\textwidth]{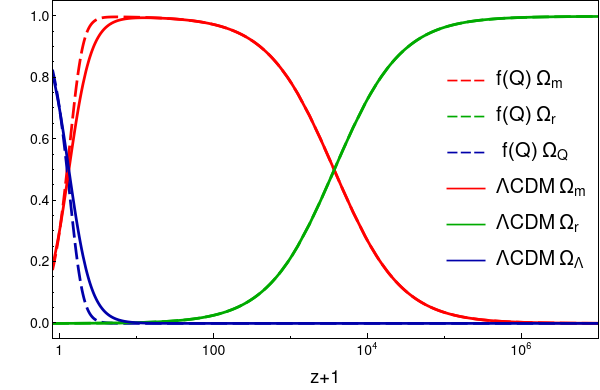}
\caption{ Evolution of the density parameters for the inverse non-metricy $f(Q)$ model  (dashed lines) and $\Lambda$CDM (solid lines) for comparison. }
\label{fig:omega_model_2}
\end{figure}

Because this model introduces no extra model parameters, there is no freedom  in fitting cosmological observables beyond what is already set by the previously defined cosmological parameters. The resulting phenomenology is therefore fully determined once those parameters are specified. We assess its impact by performing a direct comparison with $\Lambda$CDM, highlighting the signatures that emerge in the background evolution, CMB temperature (TT) angular power spectrum, the CMB lensing power spectrum, and the matter power spectrum. 

\subsection{Background evolution}
Now we can study the background evolution of the model and show it in Figure \ref{fig:omega_model_2}, where we  plot the evolution of the density parameters as a function of the redshift.   We can observe the usual succession of cosmological eras: first the dominant
contribution comes from radiation, then we have a matter dominated era and finally DE, or
in this case $Q$-fluid, takes over. By looking at Figure \ref{fig:omega_model_2} one can see that the deviations from $\Lambda$CDM only occur at lower redshifts $(z<20)$ leaving the early time cosmology the same as $\Lambda$CDM. These deviations result in a matter dominated era being longer for the inverse non-metricity model than $\Lambda$CDM.  The Q-fluid dominated era starts later than the $\Lambda$ era. The equality of $\Lambda$ and matter is $z=0.31$  while the equality of matter and Q-fluid is $z=0.24$.
 
\begin{figure}[htbp]
    \centering
\begin{subfigure}
    \centering
    \includegraphics[width=\linewidth]{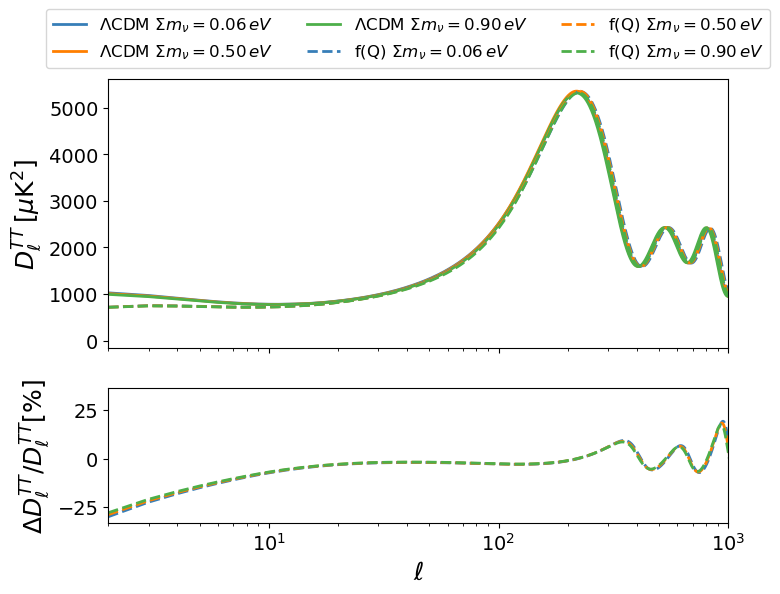}
\end{subfigure}
\begin{subfigure}
    \centering
    \includegraphics[width=\linewidth]{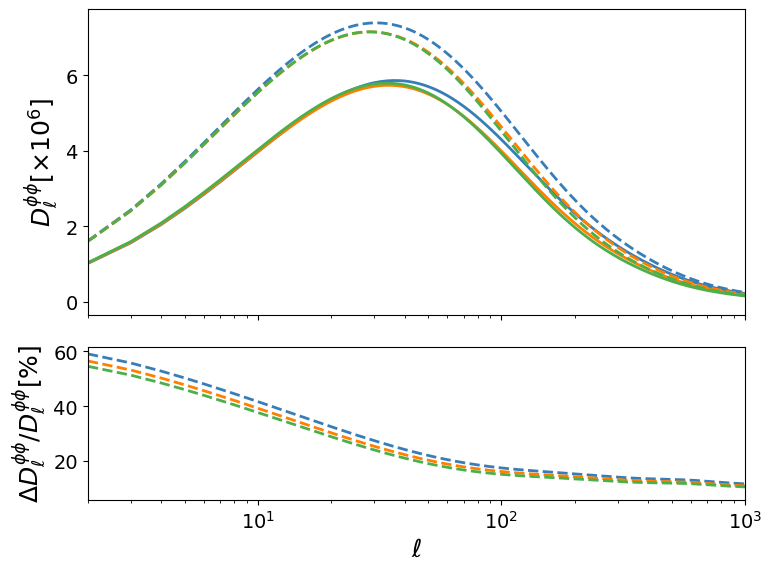}
\end{subfigure}
\begin{subfigure}
    \centering
    \includegraphics[width=\linewidth]{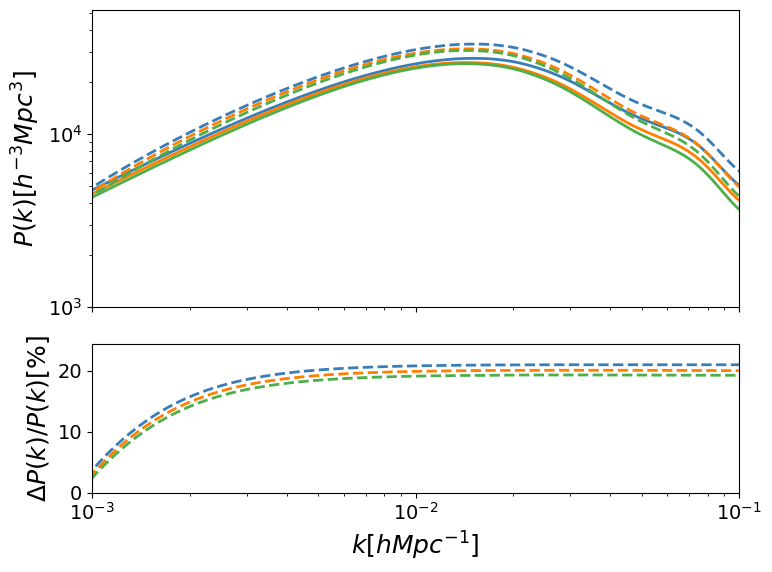}
\end{subfigure}
\caption{Top panel: CMB TT angular power spectra $D_{\ell}^{TT} = \ell(\ell+1)C_{\ell}^{TT}/(2\pi)$, middle panel: Lensing angular power spectra $D_{\ell}^{\phi \phi} = \ell^2(\ell+1)^2C_{\ell}^{\phi \phi}/(2\pi)$ and bottom panel: Matter power spectra $P(k)$ for the inverse non-metricity model and $\Lambda$CDM scenario and different values of $\Sigma m_{\nu}$. For each power spectra we present the percentage relative difference with respect to $\Lambda$CDM model.}  
\label{Phenomenology_sum_neutrinos}
\end{figure}

\subsection{Effects on cosmological power spectra}

 We assess the observational signatures of the inverse non-metricity model through a direct comparison with the $\Lambda$CDM scenario, emphasizing the differences in the CMB TT, lensing and the matter power spectra.

   For this model, we find $\mu>1$, which corresponds to a gravitational interaction that is stronger than in the standard scenario. This strengthens the growth of matter perturbations and consequently boosts the matter power spectrum with respect to $\Lambda$CDM. We note that, although the modified gravitational coupling $\mu(a)$ in QSA is scale independent, the plotted matter power spectrum is not expected to show a strictly scale-independent fractional enhancement across all 
$k$. In particular, at very large scales the spectrum is less directly sensitive to the late-time growth modification and more controlled by the primordial normalization, transfer-function physics, and gauge/horizon-scale effects. This behavior is illustrated in Figure~\ref{Phenomenology_sum_neutrinos}, bottom panels where we also show the relative deviation from $\Lambda$CDM.

A similar trend appears in the lensing sector, where the stronger gravitational interaction increases the lensing power spectrum relative to $\Lambda$CDM. This is shown in the middle panels of Figure~\ref{Phenomenology_sum_neutrinos}.

Given the modified lensing potential, variations in the integrated Sachs--Wolfe (ISW) contribution are also expected. Indeed, top panels of Figure~\ref{Phenomenology_sum_neutrinos} show a suppression of the late-time ISW contribution, leading to a reduced low-$\ell$ TT tail compared to $\Lambda$CDM \footnote{We adopt the QSA, which is formally valid on sub-horizon scales. This is a standard and controlled approximation  and has been found to work in many MG theories \cite{Sawicki:2015zya,Peirone:2017ywi,Frusciante:2018jzw}. Nevertheless, results involving horizon-scale observables should be viewed with appropriate caution.}. Moreover, the changes in the CMB TT spectrum are not limited to the ISW regime: the altered background expansion history also shifts the acoustic peak positions towards larger multipoles.

In Figure~\ref{Phenomenology_sum_neutrinos} we also show how varying the total neutrino mass $\Sigma m_{\nu}$ affects the same set of observables. In linear theory, massive neutrinos free-stream (fs) and suppress clustering below the free-streaming scale, making the growth of cold dark matter+baryon perturbations scale dependent and reducing the total matter power for $k\gtrsim k_{\rm fs}$ \cite{Lesgourgues:2006nd}.
In general, we find that  the presence of massive neutrinos suppresses the modifications induced by a stronger gravitational interaction.

\section{Data sets}\label{sec_Data}

We obtain cosmological constraints with the Markov Chain Monte Carlo sampler \texttt{MGCosmoMC}~\cite{Hojjati:2011ix} adopting the standard likelihood implementations for all datasets considered, and analyze the resulting chains using \texttt{GetDist}~\cite{Lewis:2019xzd}. In our implementation, the inverse-$f(Q)$ model modifies the background evolution and linear perturbations entering the theoretical predictions, while the observational likelihoods and nuisance-parameter treatment are otherwise kept identical to the standard pipeline.

We employ \textit{Planck} 2018~\cite{Planck:2019nip} CMB temperature data at low multipoles, $\ell\in[2,29]$, and the high-$\ell$ joint likelihoods for $TT$, $TE$, and $EE$ ($TT$: $\ell\in[30,2508]$; $TE/EE$: $\ell\in[30,1996]$). We denote this CMB combination as ``PLK18''.
We also include in their standard likelihood form BAO measurements from the SDSS DR7 Main Galaxy Sample~\cite{Ross:2014qpa} and the 6dF Galaxy Survey~\cite{Beutler_2011}, together with the combined BAO+RSD constraints from the SDSS DR12 consensus release~\cite{BOSS:2016wmc}.
In addition, we use the JLA Type Ia supernova compilation~\cite{SDSS:2014iwm} and its likelihood with the corresponding nuisance parameters marginalized over in the standard way. The joint combination of \textit{Planck}18, BAO, RSD and SNIa is referred to as ``PBRS''.

Finally, we incorporate galaxy clustering and weak-lensing measurements from DES Year-1~\cite{DES:2017myr}, including the angular two-point functions of galaxy clustering, cosmic shear, and galaxy--galaxy lensing \cite{DES:2018ufa}. The analysis includes the standard DES nuisance sector, namely galaxy-bias parameters for the clustering sample, intrinsic-alignment parameters for the shear sector, and photometric-redshift shift parameters for source and lens bins. We adopt the conservative scale cuts implemented in the likelihood, following the default DES Year-1 setup, in order to reduce sensitivity to uncertain nonlinear modelling \cite{Zucca:2019xhg,Planck:2015bue,DES:2018ufa}. We refer to the full combination including DES as ``PBRSD''.

We assume flat priors on the baseline cosmological parameters: $\Omega_b h^2\in[0.005,0.1]$, $\Omega_c h^2\in[0.001,0.99]$, $\ln(10^{10}A_s)\in[1.61,3.91]$, $\tau\in[0.01,0.8]$, $\theta_{\rm MC}\in[0.5,10]$, and $n_s\in[0.8,1.2]$.
We consider two neutrino setups: (i) a fixed minimal-mass case with $\Sigma m_\nu=0.06~\mathrm{eV}$, and (ii) a varying-mass case with $\Sigma m_\nu\in[0.01,2]~\mathrm{eV}$.

\section{Cosmological constraints}\label{sec_cosmological_constraints}

\begin{table*}[t!]
\centering
\begin{tabular}{|l|l|l|l|l|l|l|}
\hline

DATASET  & $n_s$ & $H_0 $ & $\Omega_{\rm m}^0 $ & $\sigma_8^0 $ \\ \hline \hline
\multicolumn{5}{|c|}{$\Lambda$CDM}  \\ \hline\hline
PLK18 & $0.97\pm 0.01   $ & $68.0\pm 1.4        $ & $0.31\pm 0.02   $ & $0.85\pm 0.04   $  \\
PBRS & $0.970^{+0.008}_{-0.007}$& $68.1\pm 0.8     $ & $0.30\pm 0.01   $& $0.843^{+0.032}_{-0.037}   $\\
PBRSD & $0.970\pm{0.008}$ & $68.17^{+0.77}_{-0.78} $ & $0.304^{+0.010}_{-0.0098}  $ & $0.839^{+0.030}_{-0.033}$ \\ \hline
\hline
 
 \multicolumn{5}{|c|}{ $f(Q)$-inverse power  }  \\ \hline\hline
PLK18 & $0.97\pm 0.01   $& $78.5 \pm1.7       $ &  $0.230^{+0.015}_{-0.014}   $&  $0.935^{+0.041}_{-0.040}   $\\ 

PBRS & $0.950 \pm{0.007}$ & $74.18\pm 0.86   $& $0.2703^{+0.0089}_{-0.0087}$& $0.912^{+0.024}_{-0.014}   $\\
 
PBRSD & $0.954 \pm 0.003          $& $75.21\pm 0.41             $ & $0.2596\pm 0.0040          $& $0.9008^{+0.0031}_{-0.0061}$ \\ \hline
 \end{tabular}
 \caption{Marginalized constraints on cosmological parameters at 95\% C.L. for the $\Lambda$CDM and the inverse power $f(Q)$ models with a fixed neutrino mass of 0.06 eV. }
 \label{tab:fixednu_constrains}
\end{table*}

\begin{table*}[t!]
\centering
\begin{tabular}{|l|l|l|l|l|l|l|}
\hline

DATASET  & $n_s$ & $H_0$ & $\Omega_{\rm m}^0  $ & $\sigma_8^0    $ & $\Sigma m_\nu $\\ \hline \hline
\multicolumn{6}{|c|}{$\Lambda$CDM }   \\ \hline\hline
 PLK18 & $0.97\pm0.01   $ &    $66.6^{+2.9}_{-4.1}        $ & $0.325^{+0.054}_{-0.038}   $ & $0.823^{+0.064}_{-0.080}   $ & $< 0.622                   $ \\
 PBRS  & $0.970\pm 0.008$& $67.97^{+0.87}_{-0.91}     $ & $0.306\pm 0.011   $ & $0.839\pm 0.035  $ & $< 0.219                   $ \\
 PBRSD & $0.973^{+0.010}_{-0.0093}  $& $68.08^{+0.93}_{-1.0}      $& $0.304^{+0.012}_{-0.011}   $ & $0.822^{+0.035}_{-0.034}   $& $< 0.277                   $ \\ \hline \hline
 \multicolumn{6}{|c|}{ $f(Q)$-inverse power }  \\ \hline\hline
PLK18  & $0.97\pm 0.01   $& $75.9^{+4.1}_{-5.3}  $ &  $0.251^{+0.047}_{-0.035}   $&  $0.893^{+0.075}_{-0.092}   $& $< 0.686 $\\ 

PBRS  &$0.9681 \pm 0.0095          $& $71.1^{+1.2}_{-1.11} $ & $0.294 \pm 0.011      $& $0.795 \pm 0.038$ & $0.71 \pm 0.19$\\
 
PBRSD  &$0.9695^{+0.0091}_{-0.0086}$& $71.1\pm1.1 $ & $0.292 \pm 0.011      $& $0.751 \pm 0.033$ & $0.77^{+0.18}_{-0.17}      $\\ \hline
\end{tabular}
\caption{Marginalized constraints on cosmological  parameters at 95\% C.L. for the $\Lambda$CDM and $f(Q)$ inverse power models when the total mass of the neutrinos is considered as free parameter. }
\label{Tab:VaryMassiveNeu}
\end{table*}

\begin{table*}[t!]
\centering
\begin{tabular}{|l|l|l|l|l|l|l|}
\hline
Statistics  & PLK18 & PBRS & PBRSD  \\ \hline \hline
$\Delta \chi^2=\chi^2_{f(Q)}-\chi^2_{\Lambda \rm CDM}$
  &-3.96&84.48& 124.98\\
  $\Delta \chi_{\Sigma m_\nu}^2=\chi^2_{f(Q)+\Sigma m_\nu}-\chi^2_{\Lambda \rm CDM+\Sigma m_\nu}$
  &-2.47&33.64&45.22 \\
  $\Delta {\chi^2}={\chi^2}_{f(Q)+\Sigma m_\nu}-{\chi^2}_{f(Q)}$&0.55
  &-50.69& -79.71\\
  $\Delta {\rm DIC}={\rm DIC}_{f(Q)+\Sigma m_\nu}-{\rm DIC}_{f(Q)}$&1.54
  &-50.85& -85.11
 \\ \hline
 \end{tabular}
 \caption{ Model-comparison statistics in terms of $\chi^2$ and Deviance Information Criterion (DIC). We use only $\chi^2$ statistic in the cases where the models have the same number of free parameters because information criteria reduce essentially to the same ranking as $\Delta \chi^2$. While, when comparing $f(Q)$ with varying neutrino mass with the same model but with fixed neutrino mass, we use also the DIC for comparison given that the parameters number differs.}
 \label{tab:statistics}
\end{table*}

Tables~\ref{tab:fixednu_constrains} and \ref{Tab:VaryMassiveNeu} summarize the marginalized constraints (95\% C.L.) obtained from the three dataset combinations defined in Sec.~\ref{sec_Data}: PLK18, PBRS, and PBRSD. For $\Lambda$CDM we recover the well known Planck parameter space, while the inverse non-metricity $f(Q)$ model introduces marked shifts in the late-time parameters, particularly $H_0$, $\Omega_{\rm m}^0$ and the amplitude of matter fluctuations, $\sigma_8^0$.

In $\Lambda$CDM with $\Sigma m_\nu=0.06~\mathrm{eV}$, the addition of BAO/RSD/SNIa tightens the geometric degeneracies of the CMB and reduces the uncertainties, yielding $H_0\simeq 68$ km/s/Mpc and $\Omega_{\rm m}^0\simeq 0.30$--0.31, with $\sigma_8^0\simeq 0.84$ (Table~\ref{tab:fixednu_constrains}). Including DES-1Y further improves the joint constraints without significantly shifting the posterior means.

For the inverse non-metricity $f(Q)$ model, still with $\Sigma m_\nu=0.06~\mathrm{eV}$, Planck alone prefers a substantially higher expansion rate, $H_0\simeq 78.5$ km/s/Mpc, accompanied by lower $\Omega_{\rm m}^0$ and enhanced $\sigma_8^0$ relative to $\Lambda$CDM. The PBRS combination pulls the best-fit value down to $H_0\simeq 74.2$ km/s/Mpc, while the inclusion of DES-1Y yields a tightly constrained $H_0$ and $\sigma_8^0$. This behavior is consistent with the interpretation that the MG sector can shift the background expansion to larger $H_0$, while simultaneously enhancing the growth of structure, then increasing the inferred clustering amplitude. We also note that for the $H_0$ parameter there is a slight tension between the data sets, namely CMB and BAO/RSD/SNIa, which is typical of other MG models, see e.g. \cite{Frusciante:2019puu}. This tension disappears by allowing the sum of the neutrino mass to vary.

Allowing $\Sigma m_\nu$ to vary broadens the CMB-only posteriors in $\Lambda$CDM due to the standard degeneracy between neutrino mass, $H_0$, and the late-time clustering amplitude. As expected, combining Planck with BAO/RSD/SNIa restores tight constraints and yields $H_0\simeq 68$ km/s/Mpc with an upper bound $\Sigma m_\nu \lesssim 0.2$~eV (95\% C.L.) (Table~\ref{Tab:VaryMassiveNeu}), qualitatively consistent with the well-known tightening of $\Sigma m_\nu$ limits when BAO information is included.

In the inverse non-metricity $f(Q)$ model, introducing $\Sigma m_\nu$ as a free parameter qualitatively changes the preferred region of parameter space once late-time data are included: PBRS shifts to $H_0\simeq 71.1$ km/s/Mpc and yields a comparatively large inferred neutrino mass, $\Sigma m_\nu \simeq 0.71~\mathrm{eV}$, while the inclusion of DES-1Y drives the posterior to even stronger preference for large $\Sigma m_\nu$ (Table~\ref{Tab:VaryMassiveNeu}). This trend is naturally understood as a competition between MG enhanced growth and neutrino free-streaming suppression of structure: increasing $\Sigma m_\nu$ damps the growth of matter perturbations, partially compensating the enhanced clustering sourced by the MG sector. Therefore, in this model the late-time datasets can be accommodated by moving along a degeneracy direction in which larger $\Sigma m_\nu$ offsets MG growth, resulting in a substantially reduced $\sigma_8^0$ (down to $\sim 0.75$ for PBRSD).

The $\Lambda$CDM constraints on $H_0$ are in $\sim 4$--$5\sigma$ tension with recent Cepheid-calibrated distance-ladder determinations ($H_0\simeq 73$ km/s/Mpc) \cite{Riess:2021jrx}. In contrast, the inverse non-metricity $f(Q)$ model with minimal neutrino mass prefers $H_0\simeq 74$--$78$ km/s/Mpc, moving toward the local measurements and thus reducing the $H_0$ tension. Once $\Sigma m_\nu$ is allowed to vary, the CMB only constraint on $H_0$ shows that the tension is alleviated. The preferred value shifts to an intermediate $H_0\simeq 71$ km/s/Mpc when other datasets are included; in this case the $H_0$ tension is reduced also.

\begin{figure}[t]
\centering
\includegraphics[width=0.5\textwidth]{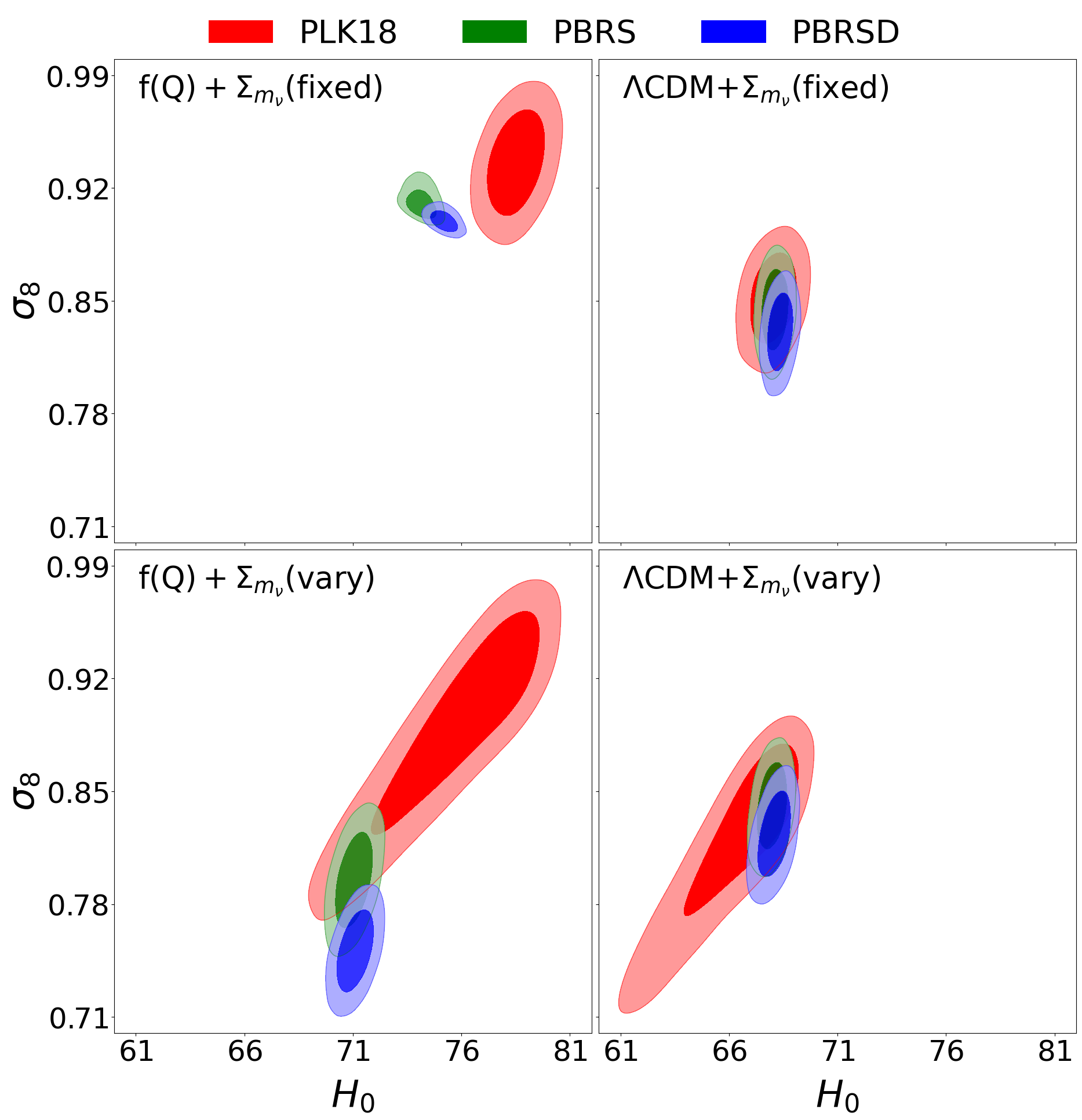}
\caption{Marginalised constraints of inverse non-metricity model and $\Lambda$CDM at 68\% (darker) and 95\% (lighter) C.L. on the $H_0$ and $\sigma_8^0$ obtained with the CMB data from Planck 2018 (PLK18, red), its combination with BAO, RSD and SNIa data (PBRS, green) and with DES-1Y data (PBRSD, blue). }
\label{fig:H_0__sigma_8__ALL}
\end{figure}

Regarding the clustering tension, weak-lensing analyses are commonly summarized by $S_8\equiv \sigma_8\sqrt{\Omega_{\rm m}/0.3}$, with typical cosmic-shear constraints in the range $S_8\simeq 0.77-0.78$. Using the values in the tables, $\Lambda$CDM yields $S_8\simeq 0.84$ for PBRSD, which lies above these weak-lensing determinations and reflects the usual $S_8$ tension. In the inverse non-metricity $f(Q)$ model with fixed minimal neutrino mass, $\sigma_8$ is enhanced while $\Omega_{\rm m}$ is reduced, leaving $S_8$ at a similarly high level; thus, the model can ease the $H_0$ tension at the expense of increased (or at best unchanged) late-time clustering tension. Allowing $\Sigma m_\nu$ to vary provides a mechanism to reduce the clustering amplitude: in the PBRSD combination we find a significantly smaller $\sigma_8$ and correspondingly lower $S_8$ (down to $\sim 0.74$), bringing the inferred clustering closer to (or below) the weak-lensing preferred values. In this scenario the simultaneous mitigation of both tensions is achieved through the interplay between MG growth enhancement and neutrino free-streaming suppression, at the cost of preferring comparatively large neutrino masses. We summarize the results about the $H_0$ and $\sigma_8$ tensions in Figure \ref{fig:H_0__sigma_8__ALL}, where we show the marginalized constraints of the $f(Q)$ model and $\Lambda$CDM with either fixed and varying  neutrino mass.  

Finally, we provide in Table \ref{tab:statistics} the statistical criteria adopted in the analysis for model comparison.
We adopt the $\chi^2$ and Deviance Information Criterion (DIC) \cite{10.1111/rssb.12062}. The $\chi^2$ statistic is used only for comparisons between models with the same number of free parameters, as in this situation other information criteria provide essentially the same ranking as $\Delta\chi^2$. In contrast, when comparing the $f(Q)$ model with free neutrino mass to the corresponding case with fixed neutrino mass, we additionally use the DIC, since the number of free parameters is no longer the same and  it allows us to take into account both the goodness of fit and the bayesian complexity of the model. Thus,  more complex models are disfavored and models with smaller DIC are preferred over models with larger DIC. For the $f(Q)$ model, allowing a higher neutrino mass improves the fit to the data for PBRSD and PBRS. This indicates a preference for larger neutrino masses within $f(Q)$. While there is no preference when CMB data only are used because in this case a larger mass of the neutrino is not allowed. This is evident for both statistical criteria adopted. When comparing $f(Q)$ to $\Lambda$CDM, the $\chi^{2}$ is only marginally lower for $f(Q)$ in the PLK18 dataset ($\Delta\chi^{2}=\chi^2_{\rm f(Q)}-\chi^2_{\rm \Lambda CDM}=-3.96$ for fixed mass of the neutrinos and $\Delta\chi_{\Sigma m_\nu}^{2}=-2.47$ when the mass of neutrino is a free parameter), while no improvement is found for the other data combinations. We conclude that the inverse 
$f(Q)$ model is strongly constrained by the combined data and it is generally not supported by data as a fully viable alternative to $\Lambda$CDM unless one allows  large neutrino masses. Overall, 
this does not provide evidence in favour of $f(Q)$ over $\Lambda$CDM.

\section{Conclusion}\label{sec_conclusion}

In this work we investigated the cosmology of an inverse non-metricity model in symmetric teleparallel $f(Q)$ gravity.
The only model parameter $M$ can be expressed in terms of the present-day matter density parameter, so that the model introduces no additional free parameters with respect to $\Lambda$CDM. 
We derived the background and linear scalar-perturbation equations in the quasi-static, sub-horizon regime, showing that departures from $\Lambda$CDM arise predominantly at late times and are governed by an effective gravitational coupling  that satisfies $\mu>1$.

We implemented the model in \texttt{MGCAMB} to obtain self-consistent predictions for CMB and large-scale-structure observables. The enhanced gravitational coupling strengthens the growth of matter perturbations, leading to an increase in the matter power spectrum and in the CMB lensing power spectrum relative to $\Lambda$CDM. At the same time, the modified late-time evolution of the gravitational potentials suppresses the late-time ISW contribution, reducing power in the low-$\ell$ CMB temperature tail, while the altered expansion history produces  shifts in the acoustic-peak positions. 

We confronted the model with current cosmological datasets using \texttt{MGCosmoMC}, considering Planck 2018 CMB data (PLK18), its combination with BAO/RSD/SNIa (PBRS), and the further inclusion of DES Year-1 galaxy clustering and weak-lensing measurements (PBRSD). 
For a fixed minimal neutrino mass, Planck-only fits favor a significantly higher $H_0$ than $\Lambda$CDM, which moves the inference toward local distance-ladder determinations and therefore eases the early--late $H_0$ tension. However, the same enhanced growth generally increases the inferred clustering amplitude, leaving the $S_8$ discrepancy unchanged or potentially exacerbated once late-time structure probes are included. 

Allowing the total neutrino mass $\Sigma m_\nu$ to vary reveals an important degeneracy: neutrino free-streaming suppresses structure growth and can partially compensate the MG enhancement. In the inverse non-metricity $f(Q)$ model, late-time data can therefore be accommodated along a direction in parameter space where larger $\Sigma m_\nu$ reduces $\sigma_8$ (and hence $S_8$), yielding values closer to weak-lensing preferences, at the cost of preferring comparatively large neutrino masses. Large neutrino masses, also eliminate the tension between CMB and BAO/RSD/SN Ia which is present with fixed (lower) neutrino mass.

From a statistical perspective, the inverse non-metricity $f(Q)$ model provides at most a marginal improvement over $\Lambda$CDM for CMB-only fits, while the addition of late-time datasets strongly constrains the available parameter space and does not yield an overall improvement in goodness-of-fit. 

These results emphasize that, despite its conceptual economy and parameter minimality, inverse $f(Q)$ gravity is highly sensitive to structure-growth and lensing measurements. Future data with improved control of late-time systematics and tighter neutrino-mass information will be particularly powerful in testing (or ruling out) this class of non-metricity models.

\acknowledgments

L.A. and S.M.N. are supported by Fundação para a Ciência e a Tecnologia (FCT) through the research grant UID/04434/2025. L.A. acknowledges support from the FCT PhD fellowship grant with ref. number 2022.11152.BD. S.M.N. acknowledges support from the FCT through the project with reference PTDC/FIS-AST/0054/2021 ("BEYond LAmbda"). N.F. acknowledge the Istituto Nazionale di Fisica Nucleare (INFN) Sez. di Napoli, Iniziativa Specifica InDark. Finally, the authors also acknowledge   the COST Action CosmoVerse, CA21136, supported by COST (European Cooperation in Science and Technology).

\bibliography{sample}

@article{Carroll:2003wy,
    author = "Carroll, Sean M. and Duvvuri, Vikram and Trodden, Mark and Turner, Michael S.",
    title = "{Is Cosmic Speed-Up Due to New Gravitational Physics?}",
    eprint = "astro-ph/0306438",
    archivePrefix = "arXiv",
    reportNumber = "FERMILAB-PUB-03-263-A, SU-GP-03-6-2",
    doi = "10.1103/PhysRevD.70.043528",
    journal = "Phys. Rev. D",
    volume = "70",
    pages = "043528",
    year = "2004"
}

@article{Frusciante:2019puu,
    author = "Frusciante, Noemi and Peirone, Simone and Atayde, Luis and De Felice, Antonio",
    title = "{Phenomenology of the generalized cubic covariant Galileon model and cosmological bounds}",
    eprint = "1912.07586",
    archivePrefix = "arXiv",
    primaryClass = "astro-ph.CO",
    reportNumber = "YITP-19-134",
    doi = "10.1103/PhysRevD.101.064001",
    journal = "Phys. Rev. D",
    volume = "101",
    number = "6",
    pages = "064001",
    year = "2020"
}

@article{Wright:2025xka,
    author = "Wright, Angus H. and others",
    title = "{KiDS-Legacy: Cosmological constraints from cosmic shear with the complete Kilo-Degree Survey}",
    eprint = "2503.19441",
    archivePrefix = "arXiv",
    primaryClass = "astro-ph.CO",
    doi = "10.1051/0004-6361/202554908",
    journal = "Astron. Astrophys.",
    volume = "703",
    pages = "A158",
    year = "2025"
}

@article{10.1111/rssb.12062,
    author = {Spiegelhalter, David J. and Best, Nicola G. and Carlin, Bradley P. and Linde, Angelika},
    title = {The Deviance Information Criterion: 12 Years on},
    journal = {Journal of the Royal Statistical Society Series B: Statistical Methodology},
    volume = {76},
    number = {3},
    pages = {485-493},
    year = {2014},
    month = {04},
    issn = {1369-7412},
    doi = {10.1111/rssb.12062},
    url = {https://doi.org/10.1111/rssb.12062},
}

@article{Peirone:2017ywi,
    author = "Peirone, Simone and Koyama, Kazuya and Pogosian, Levon and Raveri, Marco and Silvestri, Alessandra",
    title = "{Large-scale structure phenomenology of viable Horndeski theories}",
    eprint = "1712.00444",
    archivePrefix = "arXiv",
    primaryClass = "astro-ph.CO",
    doi = "10.1103/PhysRevD.97.043519",
    journal = "Phys. Rev. D",
    volume = "97",
    number = "4",
    pages = "043519",
    year = "2018"
}

@article{Frusciante:2018jzw,
    author = "Frusciante, Noemi and Peirone, Simone and Casas, Santiago and Lima, Nelson A.",
    title = "{Cosmology of surviving Horndeski theory: The road ahead}",
    eprint = "1810.10521",
    archivePrefix = "arXiv",
    primaryClass = "astro-ph.CO",
    doi = "10.1103/PhysRevD.99.063538",
    journal = "Phys. Rev. D",
    volume = "99",
    number = "6",
    pages = "063538",
    year = "2019"
}

@article{Sawicki:2015zya,
    author = "Sawicki, Ignacy and Bellini, Emilio",
    title = "{Limits of quasistatic approximation in modified-gravity cosmologies}",
    eprint = "1503.06831",
    archivePrefix = "arXiv",
    primaryClass = "astro-ph.CO",
    doi = "10.1103/PhysRevD.92.084061",
    journal = "Phys. Rev. D",
    volume = "92",
    number = "8",
    pages = "084061",
    year = "2015"
}

@article{Goncalves:2024sem,
    author = "Gon{\c{c}}alves, Tiago B. and Atayde, Lu{\'\i}s and Frusciante, Noemi",
    title = "{Cosmological study of a symmetric teleparallel gravity model}",
    eprint = "2404.01742",
    archivePrefix = "arXiv",
    primaryClass = "gr-qc",
    doi = "10.1103/PhysRevD.109.084003",
    journal = "Phys. Rev. D",
    volume = "109",
    number = "8",
    pages = "084003",
    year = "2024"
}

@article{Frusciante:2021sio,
    author = "Frusciante, Noemi",
    title = "{Signatures of $f(Q)$-gravity in cosmology}",
    eprint = "2101.09242",
    archivePrefix = "arXiv",
    primaryClass = "astro-ph.CO",
    doi = "10.1103/PhysRevD.103.044021",
    journal = "Phys. Rev. D",
    volume = "103",
    number = "4",
    pages = "044021",
    year = "2021"
}

@article{Atayde:2021pgb,
    author = "Atayde, Lu{\'\i}s and Frusciante, Noemi",
    title = "{Can $f(Q)$ gravity challenge $\Lambda$CDM?}",
    eprint = "2108.10832",
    archivePrefix = "arXiv",
    primaryClass = "astro-ph.CO",
    doi = "10.1103/PhysRevD.104.064052",
    journal = "Phys. Rev. D",
    volume = "104",
    number = "6",
    pages = "064052",
    year = "2021"
}

@article{Albuquerque:2022eac,
    author = "Albuquerque, In{\^e}s S. and Frusciante, Noemi",
    title = "{A designer approach to f(Q) gravity and cosmological implications}",
    eprint = "2202.04637",
    archivePrefix = "arXiv",
    primaryClass = "astro-ph.CO",
    doi = "10.1016/j.dark.2022.100980",
    journal = "Phys. Dark Univ.",
    volume = "35",
    pages = "100980",
    year = "2022"
}

@article{Atayde:2023aoj,
    author = "Atayde, Lu{\'\i}s and Frusciante, Noemi",
    title = "{f(Q) gravity and neutrino physics}",
    eprint = "2306.03015",
    archivePrefix = "arXiv",
    primaryClass = "astro-ph.CO",
    doi = "10.1103/PhysRevD.107.124048",
    journal = "Phys. Rev. D",
    volume = "107",
    number = "12",
    pages = "124048",
    year = "2023"
}

@article{CosmoVerseNetwork:2025alb,
    author = "Di Valentino, Eleonora and others",
    collaboration = "CosmoVerse Network",
    title = "{The CosmoVerse White Paper: Addressing observational tensions in cosmology with systematics and fundamental physics}",
    eprint = "2504.01669",
    archivePrefix = "arXiv",
    primaryClass = "astro-ph.CO",
    doi = "10.1016/j.dark.2025.101965",
    journal = "Phys. Dark Univ.",
    volume = "49",
    pages = "101965",
    year = "2025"
}

@article{Bull:2015stt,
    author = "Bull, Philip and others",
    title = "{Beyond $\Lambda$CDM: Problems, solutions, and the road ahead}",
    eprint = "1512.05356",
    archivePrefix = "arXiv",
    primaryClass = "astro-ph.CO",
    doi = "10.1016/j.dark.2016.02.001",
    journal = "Phys. Dark Univ.",
    volume = "12",
    pages = "56--99",
    year = "2016"
}

@article{Lesgourgues:2006nd,
    author = "Lesgourgues, Julien and Pastor, Sergio",
    title = "{Massive neutrinos and cosmology}",
    eprint = "astro-ph/0603494",
    archivePrefix = "arXiv",
    reportNumber = "LAPTH-1131-05, IFIC-05-59",
    doi = "10.1016/j.physrep.2006.04.001",
    journal = "Phys. Rept.",
    volume = "429",
    pages = "307--379",
    year = "2006"
}

@article{Planck:2018vyg,
  author         = "Aghanim, N. and others",
  collaboration  = "Planck",
  title          = "{Planck 2018 results. VI. Cosmological parameters}",
  journal        = "Astron. Astrophys.",
  volume         = "641",
  year           = "2020",
  pages          = "A6",
  doi            = "10.1051/0004-6361/201833910",
  eprint         = "1807.06209",
  archivePrefix  = "arXiv",
  primaryClass   = "astro-ph.CO"
}

@article{DESI:2024mwx,
  collaboration  = "DESI",
  author         = "Adame, A. G. and others",
  title          = "{DESI 2024 VI: Cosmological Constraints from the Measurements of Baryon Acoustic Oscillations}",
  journal        = "JCAP",
  volume         = "02",
  year           = "2025",
  pages          = "021",
  doi            = "10.1088/1475-7516/2025/02/021",
  eprint         = "2404.03002",
  archivePrefix  = "arXiv",
  primaryClass   = "astro-ph.CO"
}

@article{Giare:2024rgi,
  author         = "Giar\`e, William and Poulin, Vivian and Smith, Tristan L.",
  title          = "{Robust Preference for Dynamical Dark Energy in DESI BAO and SN Measurements}",
  year           = "2024",
  eprint         = "2407.16689",
  archivePrefix  = "arXiv",
  primaryClass   = "astro-ph.CO"
}

@article{Dinda:2024aag,
  author         = "Dinda, Bikash R.",
  title          = "{A new diagnostic for the null test of dynamical dark energy in light of DESI 2024 and other BAO data}",
  year           = "2024",
  eprint         = "2405.06618",
  archivePrefix  = "arXiv",
  primaryClass   = "astro-ph.CO"
}

@article{Jimenez:2017tkx,
  author         = "Jim\'enez, Jose Beltr\'an and Heisenberg, Lavinia and Koivisto, Tomi S.",
  title          = "{Coincident General Relativity}",
  journal        = "Phys. Rev. D",
  volume         = "98",
  year           = "2018",
  pages          = "044048",
  doi            = "10.1103/PhysRevD.98.044048",
  eprint         = "1710.03116",
  archivePrefix  = "arXiv",
  primaryClass   = "gr-qc"
}

@article{Heisenberg:2023uxr,
  author         = "Heisenberg, Lavinia",
  title          = "{Review on $f(Q)$ Gravity}",
  journal        = "Phys. Rept.",
  volume         = "1107",
  year           = "2024",
  pages          = "1--136",
  doi            = "10.1016/j.physrep.2024.08.001",
  eprint         = "2309.15958",
  archivePrefix  = "arXiv",
  primaryClass   = "gr-qc"
}

@article{Lazkoz:2019sji,
  author         = "L\'azkoz, Ruth and Lobo, Francisco S. N. and Ortiz-Ba\~nos, Mar\'\i a and Salzano, Vincenzo",
  title          = "{Observational constraints of $f(Q)$ gravity}",
  journal        = "Phys. Rev. D",
  volume         = "100",
  year           = "2019",
  pages          = "104027",
  doi            = "10.1103/PhysRevD.100.104027",
  eprint         = "1907.13219",
  archivePrefix  = "arXiv",
  primaryClass   = "gr-qc"
}

@inproceedings{Weinberg:2000yb,
    author = "Weinberg, Steven",
    title = "{The Cosmological constant problems}",
    booktitle = "{4th International Symposium on Sources and Detection of Dark Matter in the Universe (DM 2000)}",
    eprint = "astro-ph/0005265",
    archivePrefix = "arXiv",
    reportNumber = "UTTG-07-00",
    doi = "10.1007/978-3-662-04587-9_2",
    pages = "18--26",
    month = "2",
    year = "2000"
}

@article{Velten:2014nra,
    author = "Velten, H. E. S. and vom Marttens, R. F. and Zimdahl, W.",
    title = "{Aspects of the cosmological {\textquotedblleft}coincidence problem{\textquotedblright}}",
    eprint = "1410.2509",
    archivePrefix = "arXiv",
    primaryClass = "astro-ph.CO",
    doi = "10.1140/epjc/s10052-014-3160-4",
    journal = "Eur. Phys. J. C",
    volume = "74",
    number = "11",
    pages = "3160",
    year = "2014"
}

@article{BeltranJimenez:2019tme,
    author = "Beltr{\'a}n Jim{\'e}nez, Jose and Heisenberg, Lavinia and Koivisto, Tomi Sebastian and Pekar, Simon",
    title = "{Cosmology in $f(Q)$ geometry}",
    eprint = "1906.10027",
    archivePrefix = "arXiv",
    primaryClass = "gr-qc",
    doi = "10.1103/PhysRevD.101.103507",
    journal = "Phys. Rev. D",
    volume = "101",
    number = "10",
    pages = "103507",
    year = "2020"
}

@article{BeltranJimenez:2018vdo,
    author = "Beltr{\'a}n Jim{\'e}nez, Jose and Heisenberg, Lavinia and Koivisto, Tomi S.",
    title = "{Teleparallel Palatini theories}",
    eprint = "1803.10185",
    archivePrefix = "arXiv",
    primaryClass = "gr-qc",
    reportNumber = "NORDITA-2018-023, IFT-UAM/CSIC-18-035, IFT-UAM-CSIC-18-035",
    doi = "10.1088/1475-7516/2018/08/039",
    journal = "JCAP",
    volume = "08",
    pages = "039",
    year = "2018"
}

@article{BeltranJimenez:2017tkd,
    author = "Beltr{\'a}n Jim{\'e}nez, Jose and Heisenberg, Lavinia and Koivisto, Tomi",
    title = "{Coincident General Relativity}",
    eprint = "1710.03116",
    archivePrefix = "arXiv",
    primaryClass = "gr-qc",
    reportNumber = "NORDITA-2017-100, IFT-UAM/CSIC-17-093, ITS-ETH-2017-10",
    doi = "10.1103/PhysRevD.98.044048",
    journal = "Phys. Rev. D",
    volume = "98",
    number = "4",
    pages = "044048",
    year = "2018"
}

@article{Dialektopoulos:2019mtr,
    author = "Dialektopoulos, Konstantinos F. and Koivisto, Tomi S. and Capozziello, Salvatore",
    title = "{Noether symmetries in Symmetric Teleparallel Cosmology}",
    eprint = "1905.09019",
    archivePrefix = "arXiv",
    primaryClass = "gr-qc",
    reportNumber = "NORDITA 2019-048",
    doi = "10.1140/epjc/s10052-019-7106-8",
    journal = "Eur. Phys. J. C",
    volume = "79",
    number = "7",
    pages = "606",
    year = "2019"
}

@article{Anagnostopoulos:2021ydo,
    author = "Anagnostopoulos, Fotios K. and Basilakos, Spyros and Saridakis, Emmanuel N.",
    title = "{First evidence that non-metricity f(Q) gravity could challenge {\ensuremath{\Lambda}}CDM}",
    eprint = "2104.15123",
    archivePrefix = "arXiv",
    primaryClass = "gr-qc",
    doi = "10.1016/j.physletb.2021.136634",
    journal = "Phys. Lett. B",
    volume = "822",
    pages = "136634",
    year = "2021"
}

@article{Amendola:2007rr,
    author = "Amendola, Luca and Kunz, Martin and Sapone, Domenico",
    title = "{Measuring the dark side (with weak lensing)}",
    eprint = "0704.2421",
    archivePrefix = "arXiv",
    primaryClass = "astro-ph",
    doi = "10.1088/1475-7516/2008/04/013",
    journal = "JCAP",
    volume = "04",
    pages = "013",
    year = "2008"
}

@article{Silvestri:2013ne,
    author = "Silvestri, Alessandra and Pogosian, Levon and Buniy, Roman V.",
    title = "{Practical approach to cosmological perturbations in modified gravity}",
    eprint = "1302.1193",
    archivePrefix = "arXiv",
    primaryClass = "astro-ph.CO",
    doi = "10.1103/PhysRevD.87.104015",
    journal = "Phys. Rev. D",
    volume = "87",
    number = "10",
    pages = "104015",
    year = "2013"
}

@article{Hojjati:2011ix,
    author = "Hojjati, Alireza and Pogosian, Levon and Zhao, Gong-Bo",
    title = "{Testing gravity with CAMB and CosmoMC}",
    eprint = "1106.4543",
    archivePrefix = "arXiv",
    primaryClass = "astro-ph.CO",
    doi = "10.1088/1475-7516/2011/08/005",
    journal = "JCAP",
    volume = "08",
    pages = "005",
    year = "2011"
}

@article{Zucca:2019xhg,
    author = "Zucca, Alex and Pogosian, Levon and Silvestri, Alessandra and Zhao, Gong-Bo",
    title = "{MGCAMB with massive neutrinos and dynamical dark energy}",
    eprint = "1901.05956",
    archivePrefix = "arXiv",
    primaryClass = "astro-ph.CO",
    doi = "10.1088/1475-7516/2019/05/001",
    journal = "JCAP",
    volume = "05",
    pages = "001",
    year = "2019"
}

@article{Lewis:2019xzd,
    author = "Lewis, Antony",
    title = "{GetDist: a Python package for analysing Monte Carlo samples}",
    eprint = "1910.13970",
    archivePrefix = "arXiv",
    primaryClass = "astro-ph.IM",
    doi = "10.1088/1475-7516/2025/08/025",
    journal = "JCAP",
    volume = "08",
    pages = "025",
    year = "2025"
}

@article{Planck:2019nip,
    author = "Aghanim, N. and others",
    collaboration = "Planck",
    title = "{Planck 2018 results. V. CMB power spectra and likelihoods}",
    eprint = "1907.12875",
    archivePrefix = "arXiv",
    primaryClass = "astro-ph.CO",
    doi = "10.1051/0004-6361/201936386",
    journal = "Astron. Astrophys.",
    volume = "641",
    pages = "A5",
    year = "2020"
}

@article{Ross:2014qpa,
    author = "Ross, Ashley J. and Samushia, Lado and Howlett, Cullan and Percival, Will J. and Burden, Angela and Manera, Marc",
    title = "{The clustering of the SDSS DR7 main Galaxy sample {\textendash} I. A 4 per cent distance measure at $z = 0.15$}",
    eprint = "1409.3242",
    archivePrefix = "arXiv",
    primaryClass = "astro-ph.CO",
    doi = "10.1093/mnras/stv154",
    journal = "Mon. Not. Roy. Astron. Soc.",
    volume = "449",
    number = "1",
    pages = "835--847",
    year = "2015"
}

@article{Beutler_2011,
   title={The 6dF Galaxy Survey: baryon acoustic oscillations and the local Hubble constant: 6dFGS: BAOs and the local Hubble constant},
   volume={416},
   ISSN={0035-8711},
   url={http://dx.doi.org/10.1111/j.1365-2966.2011.19250.x},
   DOI={10.1111/j.1365-2966.2011.19250.x},
   number={4},
   journal={Monthly Notices of the Royal Astronomical Society},
   publisher={Oxford University Press (OUP)},
   author={Beutler, Florian and Blake, Chris and Colless, Matthew and Jones, D. Heath and Staveley-Smith, Lister and Campbell, Lachlan and Parker, Quentin and Saunders, Will and Watson, Fred},
   year={2011},
   month=jul, pages={3017–3032} }

@article{BOSS:2016wmc,
    author = "Alam, Shadab and others",
    collaboration = "BOSS",
    title = "{The clustering of galaxies in the completed SDSS-III Baryon Oscillation Spectroscopic Survey: cosmological analysis of the DR12 galaxy sample}",
    eprint = "1607.03155",
    archivePrefix = "arXiv",
    primaryClass = "astro-ph.CO",
    doi = "10.1093/mnras/stx721",
    journal = "Mon. Not. Roy. Astron. Soc.",
    volume = "470",
    number = "3",
    pages = "2617--2652",
    year = "2017"
}

@article{SDSS:2014iwm,
    author = "Betoule, M. and others",
    collaboration = "SDSS",
    title = "{Improved Cosmological Constraints from a Joint Analysis of the SDSS-II and SNLS Supernova Samples}",
    eprint = "1401.4064",
    archivePrefix = "arXiv",
    primaryClass = "astro-ph.CO",
    reportNumber = "FERMILAB-PUB-14-013-A-AE",
    doi = "10.1051/0004-6361/201423413",
    journal = "Astron. Astrophys.",
    volume = "568",
    pages = "A22",
    year = "2014"
}

@article{DES:2017myr,
    author = "Abbott, T. M. C. and others",
    collaboration = "DES",
    title = "{Dark Energy Survey year 1 results: Cosmological constraints from galaxy clustering and weak lensing}",
    eprint = "1708.01530",
    archivePrefix = "arXiv",
    primaryClass = "astro-ph.CO",
    reportNumber = "FERMILAB-PUB-17-294-PPD",
    doi = "10.1103/PhysRevD.98.043526",
    journal = "Phys. Rev. D",
    volume = "98",
    number = "4",
    pages = "043526",
    year = "2018"
}

@article{DES:2018ufa,
    author = "Abbott, T. M. C. and others",
    collaboration = "DES",
    title = "{Dark Energy Survey Year 1 Results: Constraints on Extended Cosmological Models from Galaxy Clustering and Weak Lensing}",
    eprint = "1810.02499",
    archivePrefix = "arXiv",
    primaryClass = "astro-ph.CO",
    reportNumber = "FERMILAB-PUB-18-507-PPD",
    doi = "10.1103/PhysRevD.99.123505",
    journal = "Phys. Rev. D",
    volume = "99",
    number = "12",
    pages = "123505",
    year = "2019"
}

@article{DES:2026fyc,
    author = "Abbott, T. M. C. and others",
    collaboration = "DES",
    title = "{Dark Energy Survey Year 6 Results: Cosmological Constraints from Galaxy Clustering and Weak Lensing}",
    eprint = "2601.14559",
    archivePrefix = "arXiv",
    primaryClass = "astro-ph.CO",
    month = "1",
    year = "2026"
}

@article{Riess:2021jrx,
    author = "Riess, Adam G. and others",
    title = "{A Comprehensive Measurement of the Local Value of the Hubble Constant with 1 km s$^{−1}$ Mpc$^{−1}$ Uncertainty from the Hubble Space Telescope and the SH0ES Team}",
    eprint = "2112.04510",
    archivePrefix = "arXiv",
    primaryClass = "astro-ph.CO",
    doi = "10.3847/2041-8213/ac5c5b",
    journal = "Astrophys. J. Lett.",
    volume = "934",
    number = "1",
    pages = "L7",
    year = "2022"
}

@article{Planck:2015bue,
    author = "Ade, P. A. R. and others",
    collaboration = "Planck",
    title = "{Planck 2015 results. XIV. Dark energy and modified gravity}",
    eprint = "1502.01590",
    archivePrefix = "arXiv",
    primaryClass = "astro-ph.CO",
    doi = "10.1051/0004-6361/201525814",
    journal = "Astron. Astrophys.",
    volume = "594",
    pages = "A14",
    year = "2016"
}

@article{Barros:2020bgg,
    author = "Barros, Bruno J. and Barreiro, Tiago and Koivisto, Tomi and Nunes, Nelson J.",
    title = "{Testing $F(Q)$ gravity with redshift space distortions}",
    eprint = "2004.07867",
    archivePrefix = "arXiv",
    primaryClass = "gr-qc",
    doi = "10.1016/j.dark.2020.100616",
    journal = "Phys. Dark Univ.",
    volume = "30",
    pages = "100616",
    year = "2020"
}

@article{Ayuso:2020dcu,
    author = "Ayuso, Ismael and Lazkoz, Ruth and Salzano, Vincenzo",
    title = "{Observational constraints on cosmological solutions of $f(Q)$ theories}",
    eprint = "2012.00046",
    archivePrefix = "arXiv",
    primaryClass = "astro-ph.CO",
    doi = "10.1103/PhysRevD.103.063505",
    journal = "Phys. Rev. D",
    volume = "103",
    number = "6",
    pages = "063505",
    year = "2021"
}

@article{Frusciante:2020gkx,
    author = "Frusciante, Noemi and Benetti, Micol",
    title = "{Cosmological constraints on Ho\v{r}ava gravity revised in light of GW170817 and GRB170817A and the degeneracy with massive neutrinos}",
    eprint = "2005.14705",
    archivePrefix = "arXiv",
    primaryClass = "astro-ph.CO",
    doi = "10.1103/PhysRevD.103.104060",
    journal = "Phys. Rev. D",
    volume = "103",
    number = "10",
    pages = "104060",
    year = "2021"
}

@article{Bellomo:2016xhl,
    author = "Bellomo, Nicola and Bellini, Emilio and Hu, Bin and Jimenez, Raul and Pena-Garay, Carlos and Verde, Licia",
    title = "{Hiding neutrino mass in modified gravity cosmologies}",
    eprint = "1612.02598",
    archivePrefix = "arXiv",
    primaryClass = "astro-ph.CO",
    doi = "10.1088/1475-7516/2017/02/043",
    journal = "JCAP",
    volume = "02",
    pages = "043",
    year = "2017"
}

@article{Motohashi:2012wc,
    author = "Motohashi, Hayato and Starobinsky, Alexei A. and Yokoyama, Jun'ichi",
    title = "{Cosmology Based on f(R) Gravity Admits 1 eV Sterile Neutrinos}",
    eprint = "1203.6828",
    archivePrefix = "arXiv",
    primaryClass = "astro-ph.CO",
    reportNumber = "RESCEU-5-12",
    doi = "10.1103/PhysRevLett.110.121302",
    journal = "Phys. Rev. Lett.",
    volume = "110",
    number = "12",
    pages = "121302",
    year = "2013"
}

@article{Dossett:2014oia,
    author = "Dossett, Jason and Hu, Bin and Parkinson, David",
    title = "{Constraining models of f(R) gravity with Planck and WiggleZ power spectrum data}",
    eprint = "1401.3980",
    archivePrefix = "arXiv",
    primaryClass = "astro-ph.CO",
    doi = "10.1088/1475-7516/2014/03/046",
    journal = "JCAP",
    volume = "03",
    pages = "046",
    year = "2014"
}

@article{He:2013qha,
    author = "He, Jian-hua",
    title = "{Weighing Neutrinos in $f(R)$ gravity}",
    eprint = "1307.4876",
    archivePrefix = "arXiv",
    primaryClass = "astro-ph.CO",
    doi = "10.1103/PhysRevD.88.103523",
    journal = "Phys. Rev. D",
    volume = "88",
    number = "10",
    pages = "103523",
    year = "2013"
}

@article{Baldi:2013iza,
    author = "Baldi, Marco and Villaescusa-Navarro, Francisco and Viel, Matteo and Puchwein, Ewald and Springel, Volker and Moscardini, Lauro",
    title = "{Cosmic degeneracies \textendash{} I. Joint N-body simulations of modified gravity and massive neutrinos}",
    eprint = "1311.2588",
    archivePrefix = "arXiv",
    primaryClass = "astro-ph.CO",
    doi = "10.1093/mnras/stu259",
    journal = "Mon. Not. Roy. Astron. Soc.",
    volume = "440",
    number = "1",
    pages = "75--88",
    year = "2014"
}

@article{Shim:2014uta,
    author = "Shim, Junsup and Lee, Jounghun and Baldi, Marco",
    title = "{Breaking the Cosmic Degeneracy between Modified Gravity and Massive Neutrinos with the Cosmic Web}",
    eprint = "1404.3639",
    archivePrefix = "arXiv",
    primaryClass = "astro-ph.CO",
    month = "4",
    year = "2014"
}

@article{Barreira:2014ija,
    author = "Barreira, Alexandre and Li, Baojiu and Baugh, Carlton and Pascoli, Silvia",
    title = "{Modified gravity with massive neutrinos as a testable alternative cosmological model}",
    eprint = "1404.1365",
    archivePrefix = "arXiv",
    primaryClass = "astro-ph.CO",
    doi = "10.1103/PhysRevD.90.023528",
    journal = "Phys. Rev. D",
    volume = "90",
    number = "2",
    pages = "023528",
    year = "2014"
}

@article{Hu:2014sea,
    author = "Hu, Bin and Raveri, Marco and Silvestri, Alessandra and Frusciante, Noemi",
    title = "{Exploring massive neutrinos in dark cosmologies with $\scriptsize{EFTCAMB}$/  EFTCosmoMC}",
    eprint = "1410.5807",
    archivePrefix = "arXiv",
    primaryClass = "astro-ph.CO",
    doi = "10.1103/PhysRevD.91.063524",
    journal = "Phys. Rev. D",
    volume = "91",
    number = "6",
    pages = "063524",
    year = "2015"
}

@article{Ballardini:2020iws,
    author = "Ballardini, Mario and Braglia, Matteo and Finelli, Fabio and Paoletti, Daniela and Starobinsky, Alexei A. and Umilt\`a, Caterina",
    title = "{Scalar-tensor theories of gravity, neutrino physics, and the $H_0$ tension}",
    eprint = "2004.14349",
    archivePrefix = "arXiv",
    primaryClass = "astro-ph.CO",
    doi = "10.1088/1475-7516/2020/10/044",
    journal = "JCAP",
    volume = "10",
    pages = "044",
    year = "2020"
}

@article{Boiza:2025xpn,
    author = "Boiza, Carlos G. and Petronikolou, Maria and Bouhmadi-L{\'o}pez, Mariam and Saridakis, Emmanuel N.",
    title = "{Addressing H $_{0}$ and S $_{8}$ tensions within f(Q) cosmology}",
    eprint = "2505.18264",
    archivePrefix = "arXiv",
    primaryClass = "astro-ph.CO",
    doi = "10.1088/1475-7516/2025/12/011",
    journal = "JCAP",
    volume = "12",
    pages = "011",
    year = "2025"
}

@article{Nashed:2025usa,
    author = "Nashed, G. G. L.",
    title = "{f(Q) gravitational theory and its structure via redshift}",
    eprint = "2502.17937",
    archivePrefix = "arXiv",
    primaryClass = "gr-qc",
    doi = "10.1140/epjc/s10052-025-13858-y",
    journal = "Eur. Phys. J. C",
    volume = "85",
    number = "2",
    pages = "183",
    year = "2025"
}

@inproceedings{Sahlu:2024pxq,
    author = "Sahlu, Shambel and Abebe, Amare",
    title = "{Constraining the modified symmetric teleparallel gravity using cosmological data}",
    eprint = "2412.20831",
    archivePrefix = "arXiv",
    primaryClass = "gr-qc",
    reportNumber = "SBN: 978-0-7961-3774-6",
    month = "12",
    year = "2024"
}

@article{Wang:2024eai,
    author = "Wang, Qingqing and Ren, Xin and Cai, Yi-Fu and Luo, Wentao and Saridakis, Emmanuel N.",
    title = "{Observational Test of f(Q) Gravity with Weak Gravitational Lensing}",
    eprint = "2406.00242",
    archivePrefix = "arXiv",
    primaryClass = "astro-ph.CO",
    doi = "10.3847/1538-4357/ad6c4d",
    journal = "Astrophys. J.",
    volume = "974",
    number = "1",
    pages = "7",
    year = "2024"
}

@article{Shi:2023kvu,
    author = "Shi, Jiaming",
    title = "{Cosmological constraints in covariant f(Q) gravity with different connections}",
    eprint = "2307.08103",
    archivePrefix = "arXiv",
    primaryClass = "gr-qc",
    doi = "10.1140/epjc/s10052-023-12139-w",
    journal = "Eur. Phys. J. C",
    volume = "83",
    number = "10",
    pages = "951",
    year = "2023"
}

@article{Ferreira:2023awf,
    author = "Ferreira, Jos{\'e} and Barreiro, Tiago and Mimoso, Jos{\'e} P. and Nunes, Nelson J.",
    title = "{Testing {\ensuremath{\Lambda}}-free f(Q) cosmology}",
    eprint = "2306.10176",
    archivePrefix = "arXiv",
    primaryClass = "astro-ph.CO",
    doi = "10.1103/PhysRevD.108.063521",
    journal = "Phys. Rev. D",
    volume = "108",
    number = "6",
    pages = "063521",
    year = "2023"
}

@article{Mandal:2021bpd,
    author = "Mandal, Sanjay and Sahoo, P. K.",
    title = "{Constraint on the equation of state parameter ($\omega$) in non-minimally coupled $f(Q)$ gravity}",
    eprint = "2111.10511",
    archivePrefix = "arXiv",
    primaryClass = "gr-qc",
    doi = "10.1016/j.physletb.2021.136786",
    journal = "Phys. Lett. B",
    volume = "823",
    pages = "136786",
    year = "2021"
}

@article{Koussour:2022zgo,
    author = "Koussour, M. and Shekh, S. H. and Hanin, A. and Sakhi, Z. and Bhoyer, S. R. and Bennai, M.",
    title = "{Flat FLRW Universe in logarithmic symmetric teleparallel gravity with observational constraints}",
    eprint = "2203.00413",
    archivePrefix = "arXiv",
    primaryClass = "gr-qc",
    doi = "10.1088/1361-6382/ac8c7d",
    journal = "Class. Quant. Grav.",
    volume = "39",
    number = "19",
    pages = "195021",
    year = "2022"
}

@article{Ferreira:2022jcd,
    author = "Ferreira, Jos{\'e} and Barreiro, Tiago and Mimoso, Jos{\'e} and Nunes, Nelson J.",
    title = "{Forecasting F(Q) cosmology with {\ensuremath{\Lambda}}CDM background using standard sirens}",
    eprint = "2203.13788",
    archivePrefix = "arXiv",
    primaryClass = "astro-ph.CO",
    doi = "10.1103/PhysRevD.105.123531",
    journal = "Phys. Rev. D",
    volume = "105",
    number = "12",
    pages = "123531",
    year = "2022"
}

@article{DAgostino:2022tdk,
    author = "D'Agostino, Rocco and Nunes, Rafael C.",
    title = "{Forecasting constraints on deviations from general relativity in f(Q) gravity with standard sirens}",
    eprint = "2210.11935",
    archivePrefix = "arXiv",
    primaryClass = "gr-qc",
    reportNumber = "ET-0236A-22",
    doi = "10.1103/PhysRevD.106.124053",
    journal = "Phys. Rev. D",
    volume = "106",
    number = "12",
    pages = "124053",
    year = "2022"
}

@article{ElBourakadi:2023sch,
    author = "El Bourakadi, K. and Sakhi, Z. and Bennai, M.",
    title = "{Observational constraints on Tachyon inflation and reheating in f(Q) gravity}",
    eprint = "2302.11229",
    archivePrefix = "arXiv",
    primaryClass = "gr-qc",
    month = "2",
    year = "2023"
}

@article{Narawade:2023rip,
    author = "Narawade, S. A. and Shekh, S. H. and Mishra, B. and Khyllep, Wompherdeiki and Dutta, Jibitesh",
    title = "{Modelling the accelerating universe with f(Q) gravity: observational consistency}",
    eprint = "2303.01985",
    archivePrefix = "arXiv",
    primaryClass = "gr-qc",
    doi = "10.1140/epjc/s10052-024-13150-5",
    journal = "Eur. Phys. J. C",
    volume = "84",
    number = "8",
    pages = "773",
    year = "2024"
}

@article{Sokoliuk:2023ccw,
    author = "Sokoliuk, Oleksii and Arora, Simran and Praharaj, Subhrat and Baransky, Alexander and Sahoo, P. K.",
    title = "{On the impact of f(Q) gravity on the large scale structure}",
    eprint = "2303.17341",
    archivePrefix = "arXiv",
    primaryClass = "astro-ph.CO",
    doi = "10.1093/mnras/stad968",
    journal = "Mon. Not. Roy. Astron. Soc.",
    volume = "522",
    number = "1",
    pages = "252--267",
    year = "2023"
}

@article{Anagnostopoulos:2022gej,
    author = "Anagnostopoulos, Fotios K. and Gakis, Viktor and Saridakis, Emmanuel N. and Basilakos, Spyros",
    title = "{New models and big bang nucleosynthesis constraints in f(Q) gravity}",
    eprint = "2205.11445",
    archivePrefix = "arXiv",
    primaryClass = "gr-qc",
    doi = "10.1140/epjc/s10052-023-11190-x",
    journal = "Eur. Phys. J. C",
    volume = "83",
    number = "1",
    pages = "58",
    year = "2023"
}

@article{Koussour:2023rly,
    author = "Koussour, M. and De, Avik",
    title = "{Observational constraints on two cosmological models of f(Q) theory}",
    eprint = "2304.11765",
    archivePrefix = "arXiv",
    primaryClass = "gr-qc",
    doi = "10.1140/epjc/s10052-023-11547-2",
    journal = "Eur. Phys. J. C",
    volume = "83",
    number = "5",
    pages = "400",
    year = "2023"
}

@article{Najera:2023wcw,
    author = "N{\'a}jera, Jos{\'e} Antonio and Alvarado, Carlos Ar{\'a}oz and Escamilla-Rivera, Celia",
    title = "{Constraints on f{\,}(Q) logarithmic model using gravitational wave standard sirens}",
    eprint = "2304.12601",
    archivePrefix = "arXiv",
    primaryClass = "gr-qc",
    doi = "10.1093/mnras/stad2180",
    journal = "Mon. Not. Roy. Astron. Soc.",
    volume = "524",
    number = "4",
    pages = "5280--5290",
    year = "2023"
}

@article{Oliveros:2023mwl,
    author = "Oliveros, A. and Acero, Mario A.",
    title = "{Cosmological dynamics and observational constraints on a viable f(Q) nonmetric gravity model}",
    eprint = "2311.01857",
    archivePrefix = "arXiv",
    primaryClass = "astro-ph.CO",
    doi = "10.1142/S0218271824500044",
    journal = "Int. J. Mod. Phys. D",
    volume = "33",
    number = "01",
    pages = "2450004",
    year = "2024"
}

@article{Sakr:2024eee,
    author = "Sakr, Ziad and Schey, Leonid",
    title = "{Investigating the Hubble tension and {\ensuremath{\sigma}} $_{8}$ discrepancy in f(Q) cosmology}",
    eprint = "2405.03627",
    archivePrefix = "arXiv",
    primaryClass = "astro-ph.CO",
    doi = "10.1088/1475-7516/2024/10/052",
    journal = "JCAP",
    volume = "10",
    pages = "052",
    year = "2024"
}

@article{YADAV2024114,
title = {Reconstructing f(Q) gravity from parameterization of the Hubble parameter and observational constraints},
journal = {Journal of High Energy Astrophysics},
volume = {43},
pages = {114-125},
year = {2024},
issn = {2214-4048},
doi = {https://doi.org/10.1016/j.jheap.2024.06.012},
url = {https://www.sciencedirect.com/science/article/pii/S2214404824000557},
author = {Anil Kumar Yadav and S.R. Bhoyar and M.C. Dhabe and S.H. Shekh and Nafis Ahmad}
}

@article{Pradhan:2024eew,
    author = "Pradhan, Sneha and Solanki, Raja and Sahoo, P. K.",
    title = "{Cosmological constraints on f(Q) gravity models in the non-coincident formalism}",
    eprint = "2410.00922",
    archivePrefix = "arXiv",
    primaryClass = "gr-qc",
    doi = "10.1016/j.jheap.2024.08.002",
    journal = "JHEAp",
    volume = "43",
    pages = "258--267",
    year = "2024"
}

@article{Sharma:2024gbl,
    author = "Sharma, Lokesh Kumar and Parekh, Suresh and Yadav, Anil Kumar",
    title = "{Observational constraints using Bayesian statistics and deep learning in f(Q) gravity}",
    eprint = "2412.12323",
    archivePrefix = "arXiv",
    primaryClass = "gr-qc",
    doi = "10.1016/j.nuclphysb.2025.117007",
    journal = "Nucl. Phys. B",
    volume = "1018",
    pages = "117007",
    year = "2025"
}

@article{Oliveira:2025qvn,
    author = "Oliveira, Fernanda and Ribeiro, Bruno and Hip{\'o}lito-Ricaldi, Wiliam S. and Avila, Felipe and Bernui, Armando",
    title = "{Viability of general relativity and modified gravity cosmologies using high-redshift cosmic probes}",
    eprint = "2505.19960",
    archivePrefix = "arXiv",
    primaryClass = "astro-ph.CO",
    doi = "10.1088/1475-7516/2025/12/007",
    journal = "JCAP",
    volume = "12",
    pages = "007",
    year = "2025"
}

@ARTICLE{2025EPJC...85..656K,
       author = {{Kolhatkar}, Ameya and {Mishra}, Sai Swagat and {Sahoo}, P.~K.},
        title = "{Implications of cosmological perturbations of Q in STEGR}",
      journal = {European Physical Journal C},
         year = 2025,
        month = jun,
       volume = {85},
       number = {6},
          eid = {656},
        pages = {656},
          doi = {10.1140/epjc/s10052-025-14384-7},
       adsurl = {https://ui.adsabs.harvard.edu/abs/2025EPJC...85..656K}
}

@article{Sultanaa:2025ooz,
    author = "Sultanaa, Sanjeeda and Chattopadhyay, Surajit",
    title = "{Constraining exponential f(Q) gravity with cosmic chronometers and Supernovae: A data-driven analysis}",
    eprint = "2511.06088",
    archivePrefix = "arXiv",
    primaryClass = "gr-qc",
    doi = "10.1016/j.jheap.2025.100422",
    journal = "JHEAp",
    volume = "48",
    pages = "100422",
    year = "2025"
}

@article{Karmakar:2025yng,
    author = "Karmakar, Purnendu and Haridasu, Sandeep",
    title = "{Dynamical Dark Energy or Modified Gravity? Signatures in Gravitational Wave Propagation}",
    eprint = "2509.07976",
    archivePrefix = "arXiv",
    primaryClass = "gr-qc",
    month = "9",
    year = "2025"
}

@article{Li:2025msm,
    author = "Li, Chunyu and Ren, Xin and Yang, Yuhang and Saridakis, Emmanuel N. and Cai, Yi-Fu",
    title = "{Decoupling perturbations from background in $f(Q)$ gravity: the square-root correction and the alleviation of the $σ_8$ tension}",
    eprint = "2512.16551",
    archivePrefix = "arXiv",
    primaryClass = "astro-ph.CO",
    month = "12",
    year = "2025"
}

@article{Aggarwal:2022eae,
    author = "Aggarwal, Nakul and Pourmand, Ali and Shojai, Fatimah and Parthasarathy, Harish",
    title = "{Constraining Generalized Chaplygin Gas in Non-Minimally Coupled $f(Q)$ Cosmology using Quasars and $H(z)$ Data}",
    eprint = "2212.00312",
    archivePrefix = "arXiv",
    primaryClass = "gr-qc",
    month = "12",
    year = "2022"
}

\end{document}